\documentclass[pra,twocolumn]{revtex4}%
\usepackage{amsfonts}
\usepackage{amsmath}
\usepackage{amssymb}
\usepackage{graphicx}%
\providecommand{\U}[1]{\protect\rule{.1in}{.1in}}
\begin{document}
\title{Heating and Cooling in Adiabatic Mixing process}
\author{Jing Zhou$^{1}$}
\author{Zi Cai$^{2}$}
\author{Xu-Bo Zou$^{1}$}
\email{xbz@ustc.edu.cn}
\author{Guang-Can Guo$^{1}$}
\affiliation{$^{1}$Key Laboratory of Quantum Information, Department of Physics, University
of Science and Technology of China, Hefei 230026, People's Republic of China}
\affiliation{$^{2}$Beijing National Laboratory for Condensed Matter Physics, Institute of
Physics, Chinese Academy of Sciences, Beijing 100080, P. R. China}

\begin{abstract}
We study the effect of interaction on the temperature change in the process of
adiabatic mixing two components of fermi gases by the real-space Bogoliubov-de
Gennes (BdG) method. We find that in the process of adiabatic mixing, the
competition of the adiabatic expansion and the attractive interaction make it
possible to cool or heal the system depending on strength of interactions and
the initial temperature. The change of temperature in a bulk system and a
trapped system have been investigated respectively.

\end{abstract}

\maketitle

%advantage of highly controllability

Exciting development in ultracold atom systems has opened a new possibility to
stimulate the many-body Hamiltonian that have been used to study strongly
correlated systems in condensed matter physics. Among the most exciting
breakthroughs are experimental realization of the quantum phase transition
from the superfluid to Mott-insulating phase in bosonic system \cite{Greiner},
and the metal-insulator transition with fermionic atoms
\cite{Jordens,bloch_mi} in optical lattice. Many of these many-body phenomena
in ultracold atom systems are sensitive to temperature, for an example, a
change in temperature due to adiabatic or non-adiabatic tuning of the
parameters in the many-body Hamiltonian may hide the signature of the quantum
phase transition, replacing it with a thermal transition instead. Adiabatic
process, which keeps the entropy of the system a constant, is known to play an
important role in experimental manipulation of the ultracold atoms, especially
in cooling the many-body systems. Therefore, how temperature changes in the
process of adiabatic tuning the many-body Hamiltonian is a question of great
interest and has important practical application in the experiment due to its
potential relation with cooling the ultracold atoms. In this process, the
interaction have been known to play a key role in determining the temperature
changing \cite{Ho,Pollet,Blakie1,Blakie2,Ho1,Cramer,Werner,Dare,Jean,Kurn},
and may lead to anomalous phenomenons \cite{Lucia}.

In this paper, we study the effect of interaction on the temperature change in
the process of adiabatic mixing two components of fermi gases. Initially, the
spin$\uparrow$ and spin$\downarrow$ fermions are well separated, and the gases
are completely mixed finally. We introduce the s-wave interaction between the
spin up and down fermions, which could be adjusted by s-wave Feschbach
resonance\cite{a1,a2}. We assume the interaction is attractive thus we can
safely use real-space BdG method to analyze this question. Without the
interaction, the adiabatic mixing of the two component fermions would
definitely cool the system. Below we focus on the effect of the interaction.
Whether the process of the adiabatic mixing in this interacting many-body
system would cool or heating the system? We would show below that the answer
to this question not only depends on the strength of the interaction, but also
on the initial temperature of our system.

\begin{figure}[ptb]
\includegraphics[width=8.5cm]{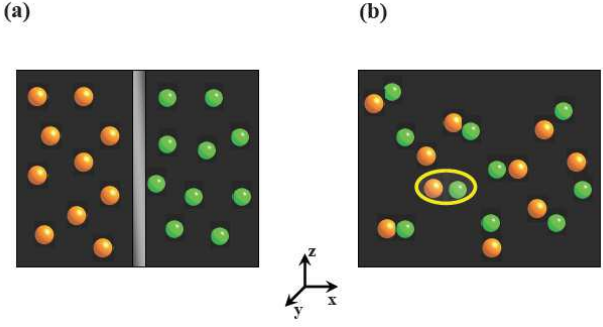} \newline\caption{Two configurations
of two component fermi gases in homogenous bulk system. (a)Separate
noninteracting fermi gases; (b)Fully-mixed fermonic superfluidity.}%
\label{111}%
\end{figure}

The paper is organized as follows: firstly, we study the thermodynamic
properties of an ideal situation and calculate the change of temperature from
an ideal initial configurations (a): a completely separate fermi gases to an
ideal final configuration (b): a mixture of attractive fermions (as shown in
Fig.\ref{111}). Further more, we propose an experimental setup to realize this
adiabatic mixing process by introducing a gradient magnetic field, which
recently been applied as a super cool atom thermometer\cite{Weld}. We show
that by tuning the gradient of the magnetic field from a large value to zero
(as shown in Fig.\ref{MM}) slowly enough, Configuration (a) and (b) in
Fig.\ref{111} can be connected adiabatically. Using the real space
self-consistent Bogoliubov-de Gennes (BdG) method, we study the temperature
change following the isoentrope and observe the variation of the real-space
distribution of the particle number as well as the superconductor order
parameters in the process of adiabatic mixing. In the final part of the paper,
we briefly discuss the possible application and relations of our result on
recent experiments about the mixing of two-component fermi gases.

Firstly, we study an ideal situation. Considering two situations: one is the
two-component fermions separated completely and there is no interaction
between them, as shown in Fig.\ref{111}(a); The other is fermi gases fully
mixed and the attractive interaction between spin up and spin down fermions
lead to pairing between them, as shown in Fig.\ref{111}(b). Suppose there is a
isoentropic process from Configuration (a) to Configuration (b), we address
the question how do the temperature change? In the isoentropic process, it is
well known that the expansion of the fermions would cool the
system\cite{Jean,Kurn}, however, the mixing of attractive spin up and spin
down fermions leads to pairing between them, which would heat the system
considering the entropy of the fermionic superfluidity is lower than the
normal state in the same temperature. The temperature change obviously depends
on the strength of the interaction since we are dealing with a many-body
systems. Further more, we find that it also depends on the initial temperature
of the well-separated noninteracting state. In the region of very low
temperature, the pairing effect on the temperature change overwhelm that of
the adiabatic expansion thus the temperature of the system would increase in
this isoentropic process. When the temperature is close to the critical
temperature $T_{c}$, the pairing effect is not important and the temperature
change mainly determined by the adiabatic mixing, which leads to the cooling
of the system. 

For situation (a), the entropy and particle number at finite temperature in
the non interacting fermionic gases are: are as follows:%
\begin{align}
N &  =N_{\uparrow}+N_{\downarrow}=\sum_{\mathbf{k}}2k_{B}f(\epsilon
_{\mathbf{k}}-\mu_{a},T)\\
S_{a} &  =-2k_{B}\sum_{\mathbf{k}}[f(\epsilon_{\mathbf{k}}-\mu_{a},T)\ln
f(\epsilon_{\mathbf{k}}-\mu_{a}),T)\nonumber\\
&  +f(-(\epsilon_{\mathbf{k}}-\mu_{a}),T)\ln f(-(\epsilon_{\mathbf{k}}-\mu
_{a}),T)]
\end{align}
while for situation (b), the corresponding thermodynamic properties are:%
\begin{align}
N &  ={\displaystyle\sum\limits_{\mathbf{k}}}(1-\frac{\epsilon_{\mathbf{k}%
}-\mu_{b}}{E_{\mathbf{k}}}\tanh\frac{E_{\mathbf{k}}}{2k_{B}T})\\
S_{b} &  =-2k_{B}{\displaystyle\sum\limits_{\mathbf{k}}}[f(E_{\mathbf{k}%
},T)\ln f(E_{\mathbf{k}},T)\nonumber\\
&  +f(-E_{\mathbf{k}},T)\ln f(-E_{\mathbf{k}},T)]\\
\text{ \ \ }\frac{m}{4\pi\hbar^{2}a} &  =\sum\limits_{\mathbf{k}}\frac{1}%
{2}(\frac{\tanh\frac{E_{\mathbf{k}}}{2k_{B}T}}{E_{\mathbf{k}}}-\frac
{1}{\epsilon_{\mathbf{k}}})
\end{align}
where $\mu_{i}$ is chemical potential and $T$ is temperature, while $S$ is the
entropy of the system and $N$ is the total particle number. $f(x,T)$ is the
function of fermi distribution at temperature $T$. $a$ is the scattering
length of attractive interaction. $\epsilon_{\mathbf{k}}=\frac{\hbar^{2}k^{2}}{2m},%
E_{k}=\sqrt{(\epsilon_{\mathbf{k}}-\mu_{b})^{2}+\Delta^{2}},$ where
$\Delta$ is the pairing parameter for BCS state. The fermi momentums of free
particle in two above situations have a relationship $\frac{k_{Fa}^{3}}%
{2}=k_{Fb}^{3}=k_{F}^{3}$. In our calculation, $k_{F}$ and $\epsilon_{F}$ is
used as scaling quantity of momentum and energy, respectively. The curves of
entropy vs temperature of such two situations are shown in Fig. 2(a).
Obviously, the entropy of both situations must be extreme low near $T=0$
according to the third law of thermodynamics.

\begin{figure}[ptb]
\includegraphics[width=7.7cm]{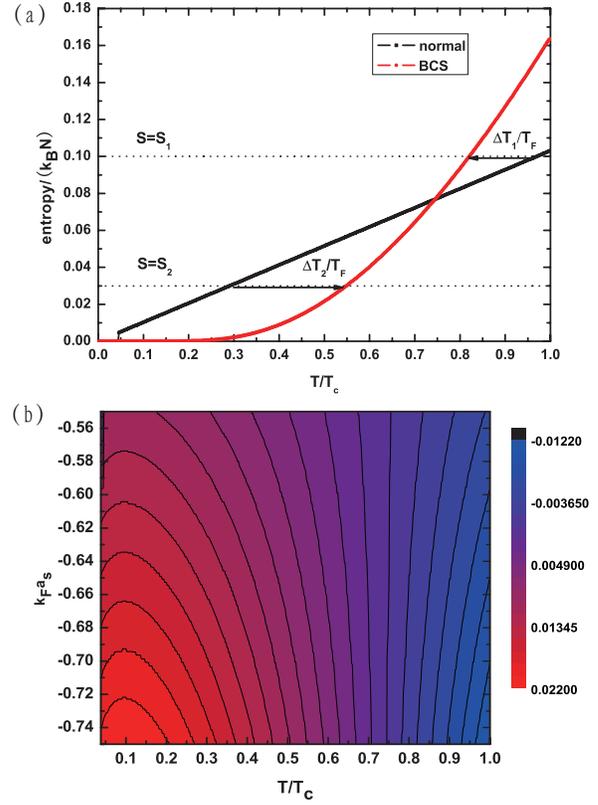}\newline\caption{(a) S-T curves in
the homogenous bulk system for separated noninteracting fermi gases and fully
mixed fermionic superfluidity. The interaction parameter is set as $k_{f}%
a_{s}=-0.6$. Two dashed lines denote two isoentropic processes. $\Delta
T_{i}/T_{F}$ is the variation of temperature during the adiabatic process
(blue is cooling while red is heating process). (b)Dependence of the
temperature change in the isoentropic process on the initial temperature $T$
as well as the strength of the interaction.}%
\end{figure}

We find fermonic superfluidity (BCS state) has the lower entropy while the
temperature is not high enough. However, when we increase the temperature, the
long-range order is destroyed gradually by the thermal fluctuation. When
temperature has reaches to $0.75T_{c}$, the entropy curves of attractive gases
and that of completely separated gases intersect. If the temperature continues
to rise, the fermi gases with attractive interaction change closer to normal
mixture, of which the entropy is certainly larger than that of the separated
gas, as shown in our result. Cooper pairs in BCS state are completely
destroyed at the critical temperature $T_{c}$ of phase transition. The curves
of entropy has shown the competition between quantum order and thermal
fluctuation. We assume the separated fermi gases can be mixed adiabatically.
The dashed lines in Fig. 2(a) denote two isoentropic processes with entropy
fixed at $S_{1}$ and $S_{2}$, and the first one is cooling while the seconde
is heating. Fig. 2(b) has shown the change of temperature in the adiabatic
processes with different interaction strength, from which we can clearly see
the heating and cooling regions respectively.

%\section{trapped system}

Now we discuss the experimental realization of this adiabatic mixing process
by applying a gradient magnetic field, which was recently used as a super cool
atom thermometer\cite{Weld}. The gradient of the magnetic field is along the
x-direction $B(\mathbf{x})=B(x)\hat{\mathbf{z}}$. By tuning the gradient of
the magnetic field from a large value to zero , configuration (a) and (b) in
Fig.\ref{MM} are connected adiabatically, as shown in Fig.\ref{MM}. we assume
that there is a hard constraint along x-direction while fermions are free
along $y$ and $z$ directions :%
\begin{equation}
V_{trap}(\mathbf{x})=\left\{
\begin{array}
[c]{ll}%
\infty, & \hbox{($x\subset[-L/2,L/2]$);}\\
0, & \hbox{(others).}
\end{array}
\right.
\end{equation}
\begin{figure}[ptb]
\includegraphics[width=8.5cm]{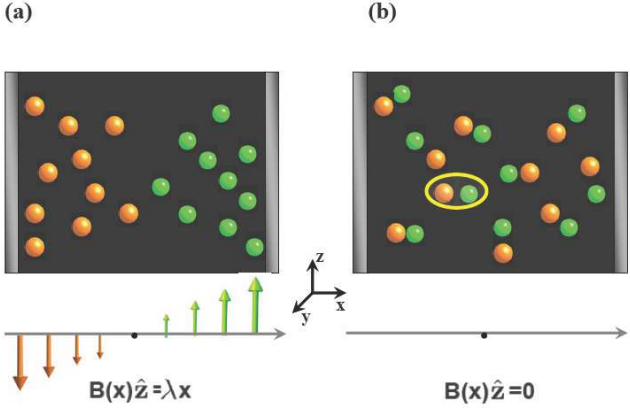}\newline\caption{(a) For the magnetic
field with large gradient, fermions with different spin are well separated;
(b) By adiabatically tuning the gradient of the magnetic field to zero, the
fermions are fully mixed.}%
\label{MM}%
\end{figure}To determine the thermodynamic properties such as the entropy and
distribution of the particles in our system, we use the real-space BdG method
to study this fermionic superfluidity under this gradient magnetic field. The
hamiltonian in this case can be written as:
\begin{align}
H  &  =\sum_{\sigma}\int d\mathbf{x\Psi}_{\sigma}^{\dag}(\mathbf{x}%
)(\frac{-h^{2}}{2m}\triangledown^{2}-\mu_{\sigma}+V_{\sigma}(\mathbf{x}%
))\mathbf{\Psi}_{\sigma}(\mathbf{x})\nonumber\\
&  +g\int d\mathbf{x\Psi}_{\uparrow}^{\dag}(\mathbf{x})\mathbf{\Psi
}_{\downarrow}^{\dag}(\mathbf{x})\mathbf{\Psi}_{\downarrow}(\mathbf{x}%
)\mathbf{\Psi}_{\uparrow}(\mathbf{x}) \tag{4}%
\end{align}
$\mu_{\sigma}$ is the chemical potential of spin $\sigma$. $V_{\sigma
}(\mathbf{x})$ include trap potential $V_{trap}$ and Zeeman shift $V_{\sigma
}^{Zee}$. Here, we assume Zeeman shift of two components has form as
$V_{\sigma}^{Zee}(\mathbf{x})=\sigma\widetilde{\lambda}x$ $(\widetilde
{\lambda}>0)$. Due to opposite energy shift, fermions with different spin are
pulled to opposite direction along x axis. Since there is not extra gradient
potential and constraint along y,z directions, which means the physical
quantities in our systems such as density, fermionic pairing order parameter
are uniform along $y$ and $z$ directions, thus they are only the function of
$x$ in real space.

\begin{figure}[ptb]
\includegraphics[scale=0.45]{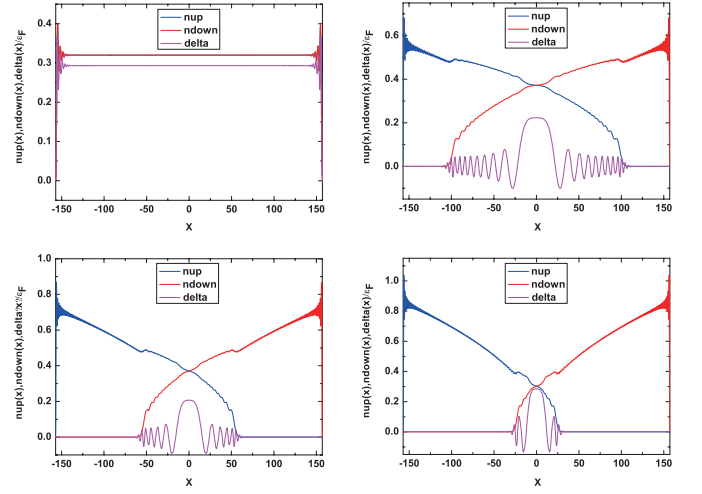}\newline%
\caption{Distribution of density and pairing parameter. The total number of
particle $N=200$ and the bare interaction parameter $g$ is set as -1.937. We
choose the temperature $T=2\times10^{-6}T_{F}$, and (a)$\widetilde{\lambda}%
$=0, (b)$\widetilde{\lambda}$=0.016, (c)$\widetilde{\lambda}$=0.0286,
(d)$\widetilde{\lambda}$=0.0446.}%
\label{a}%
\end{figure}

In order to show the density distribution in this inhomogenous system, we
adopt to solve BdG equations in real space \cite{hu1,hu2,torma}. Within
mean-field approximation, the pairing gap and density are defined as
$\triangle(x)=-g\langle\Psi_{\downarrow}(x)\Psi_{\uparrow}(x)\rangle$, where g
is the bare interaction parameter, $n_{\sigma}(x)=\langle\Psi_{\sigma
}^{\dagger}(x)\Psi_{\sigma}(x)\rangle$. We use Bogoliubov transformation
$\Psi_{\sigma}(x)=\sum_{l}(u_{l\sigma}(x)c_{l\sigma}+\sigma v_{l\overline
{\sigma}}^{\ast}c_{l\overline{\sigma}})$ to diagonalize the hamiltonian and
get BdG equation:
\begin{equation}
\left[
\begin{array}
[c]{cc}%
H_{\sigma}-\mu_{\sigma} & \Delta(x)\\
\Delta(x)^{\ast} & -H_{\overline{\sigma}}+\mu_{\overline{\sigma}}%
\end{array}
\right]  \left[
\begin{array}
[c]{c}%
u_{l\sigma}\\
v_{l\sigma}%
\end{array}
\right]  =E_{l\sigma}\left[
\begin{array}
[c]{c}%
u_{l\sigma}\\
v_{l\sigma}%
\end{array}
\right]  \tag{5}%
\end{equation}
Here, $H_{\sigma}=\frac{-\hbar^{2}}{2m}\nabla^{2}+gn_{\sigma}(x)+\sigma
\widetilde{\lambda}x$ and $l$ denotes different eigenstate of quasi particle.
Since there is some symmetry between spin up and down, $\left[
\begin{array}
[c]{c}%
u_{l\uparrow}(x)\\
v_{l\uparrow}(x)
\end{array}
\right]  =\left[
\begin{array}
[c]{c}%
-v_{l\overline{\sigma}}^{\ast}(x)\\
u_{l\overline{\sigma}}^{\ast}(x)
\end{array}
\right]  $ and $E_{l\uparrow}=-E_{l\downarrow}$, we can drop $\sigma$ in the
above matrix equation. Pairing gap and density can be expressed by the quasi
particle wave function $u_{l}(x)$ and $v_{l}(x)$: $\Delta(x)=g\sum_{l}%
u_{l}(x)v_{l}(x)^{\ast}f(E_{l}),$ $n_{\uparrow}(x)=\sum_{l}\left\vert
u_{l}(x)\right\vert ^{2}f(E_{l})$ and $n_{\downarrow}(x)=\sum_{l}\left\vert
v_{l}(x)\right\vert ^{2}f(-E_{l}),$ where $f(x)$ is Fermi-Dirac function at
temperature $T$. We have adopt the hybrid process introduced in \cite{hu1,hu2}%
. Besides the discrete spectra, the high energy part above an enough energy
cutoff is included by local density approximation. The bare interaction $g$ is
replaced by a position-dependent effective interaction $g_{eff}(x).$ Since the
particle number of each component should conserve, we obtain the distribution
of pairing parameter $\Delta(x)$ and density by self-consistent calculation.
As shown in Fig. 4, where we can find fermions with different spin are
separated gradually while the scale and amplitude of $\Delta(x)$ decreases
rapidly with the increasing of magnetic gradient. \begin{figure}[ptb]
\includegraphics[width=7.0cm]{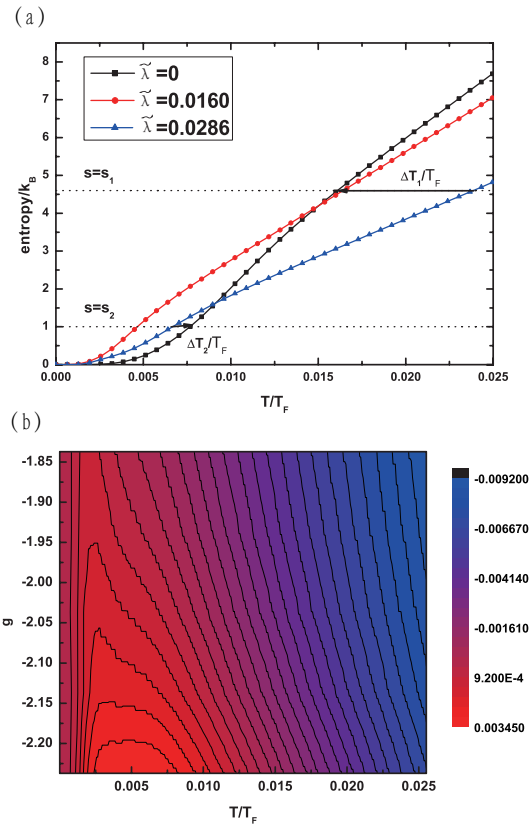}\newline\caption{(a) S-T
curves in the trapped system with different magnetic field gradient
$\widetilde{\lambda}=0,0.016,0.0286$. The bare interaction parameter $g$ is
set as -1.937. (b) contour of temperature change of the trapped system while
the magnetic field gradient is tuned from $\widetilde{\lambda}=0.0286$ to
$\widetilde{\lambda}=0$ adiabatically. }%
\label{a}%
\end{figure}

Now we return to address the same question proposed above: how do the
temperature change in this adiabatic process. By using the equation of entropy
$S=-k_{B}\sum_{l}[f(E_{l})\ln f(E_{l})+f(-E_{l})\ln f(-E_{l})]$, we can
calculate the physical quantities and we can draw the curves of entropy vs
temperature with different magnetic gradient $\widetilde{\lambda
}=0,0.016,0.0286$. (Notice that Fig.1(a) corresponding two extreme condition
of this case) Heating and cooling during the isoentropic processes are denoted
by two dashed lines in the figuration. The change of temperature in the
adiabatic mixing process as a function of initial state temperature $T$ and
strength of interaction $g$ is shown in Fig. 5(b), where we set the gradient
of the magnetic field is tuned from $\widetilde{\lambda}=0.0286$ to
$\widetilde{\lambda}=0$ adiabatically. As shown in the figure, both healing
and cooling occur in our case.

%\section{Conclusion and Discussion}
In conclusion, we have shown the change of temperature for attractive fermi
gases during the adiabatic mixing process in a homogenous bulk system and a
trapped system. The competition between the effect of the adiabatic mixing and
the interaction on the entropy lead to interesting cooling and healing
process. We focus on the attractive case where we can safely use mean-field
method. Recently, the itinerant ferromagnetism has been discovered in a
two-component fermi gases with repulsive interactions \cite{ketterlt}. More
recently, the mixing of two spin components of a strongly interacting Fermi
gas have been realized experimentally \cite{Sommer}. However, for the
repulsive interaction, the mean-field result may be unreliable and the similar
question how do the temperature change in the process of adiabatic mixing of
fermi gases with repulsive interaction may be more interesting and deserve
further investigation via other numerical method such as dynamic mean-field
theory (DMFT)\cite{DMFT}.

\textbf{Acknowledgments:} This work was funded by National Natural Science
Foundation of China (Grant No. 10574022 and Grant No. 60878059) and
\textquotedblleft Hundreds of Talents \textquotedblright\ program of the
Chinese Academy of Sciences.

\end{document}